# A Study of Selectively Digital Etching Silicon-Germanium with Nitric and Hydrofluoric Acids

*Chen Li[†, ‡], Huilong Zhu[†, ‡, §], Yongkui Zhang[†], Xiaogen Yin[†, ‡], Kunpeng Jia[†], Junjie Li[†], Guilei Wang[†], Zhenzhen Kong[†], Anyan Du[†], Tengzhi Yang[†, ‡], Liheng Zhao[†, §], Lu Xie[†, ‡], Xuezheng Ai[†], Shishuai Ma[†], Yangyang Li[†, ‡], Henry H. Radamson[†, ‡]*

[†]Key Laboratory of Microelectronics Devices & Integrated Technology, Institute of Microelectronics, Chinese Academy of Sciences, Beijing 100029, China.

[‡]University of Chinese Academy of Sciences, Beijing 100049, China.

[§]University of Science and Technology of China, Hefei, Anhui 230026, People's Republic of China.

ABSTRACT: A digital etching method was proposed to achieve excellent control of etching depth. The digital etching characteristics of p+ Si and $Si_{0.7}Ge_{0.3}$ using the combinations of $HNO_3$ oxidation and BOE oxide removal processes were studied. Experiments showed that oxidation saturates with time due to low activation energy. A physical model was presented to describe the wet oxidation process with nitric acid. The model was calibrated with experimental data and the oxidation saturation time, final oxide thickness, and selectivity between $Si_{0.7}Ge_{0.3}$ and p+ Si were obtained. The digital etch of laminated $Si_{0.7}Ge_{0.3}$/p+ Si was also investigated. The depth of the tunnels formed by etching SiGe layers between two Si layers was found in proportion to digital etching cycles. And oxidation would also saturate and the saturated relative etched amount per cycle (REPC) was 0.5 nm (4 monolayers). A corrected selectivity calculation formula was presented. The oxidation model was also calibrated with $Si_{0.7}Ge_{0.3}$/p+ Si stacks, and selectivity from model was the same with the corrected formula. The model can also be used to analyze process variations and repeatability. And it could act as a guidance for experiment design. Selectivity and repeatability should make a trade-off.

1. INTRODUCTION

SiGe is a promising material to replace the Si channel owing to high electron and/or hole mobilities, high density-of-states avoiding source starvation, high compatibility with the mainstream Si-based processes. SiGe/Si laminated structures have been used to fabricate nanowires/nanosheets (NWs/NSs) at advanced 3 nm technology node[1-2]. However, difficulties exist in gate definition and channel size control[3-4] due to high selective Si nanowire release and inner spacer cavity formation of 5~10 nm[5-7]. To solve these problems, selective etching with accurate etching depth control and high selectivity was essential. Several selective etching methods have been reported, such as dry etchings with $HCl$[8] or mixtures of $CF_4/O_2/N_2$[9-10] $CF_4/O_2/He$[11] and wet etchings with mixtures of $HNO_3$, $HF$, $H_2O$ (HNA)[12] or $H_2O_2$, $HF$, $CH_3COOH$[13-14]. However, all these methods are continuous etching and etching depth is time-dependent. They will bring bad repeatability while achieving high selectivity. Besides, dry etching exists loading effect[15-16], HNA system is sensitive to defects[17-19]. Digital etching or atomic layer etching (ALE) method is proposed to provide great control of etch variability because of its self-limiting characteristic and low etch rate[20-24]. However, to our knowledge, few reports provide the selectively digital etching characteristic of laminated structure, such as Si/SiGe. The digital etching of SiGe and Si has not been analyzed systematically.

In this work, the characteristics of Si/SiGe digital etching were studied in flat surfaces of p+ Si and $Si_{0.7}Ge_{0.3}$ using $HNO_3$ oxidation and BOE removal processes. $Si_{0.7}Ge_{0.3}$ and p+ Si oxidization saturation in $HNO_3$ and the saturation time were found. A physical model for the oxidation process was presented and calibrated with our experiments. Then the etching of laminated p+ $Si/Si_{0.7}Ge_{0.3}$ was investigated and calibrated with our oxidation model. The selectivity was calculated with our corrected method. Different etching characteristic exists between flat and laminated structures.

## 2. THE SiGe SELECTIVELY DIGITAL ETCHING METHOD

In the 1960s, Robbins and Schwartz are the first who studied systematically the etching theory of silicon in the $HNO_3/HF/H_2O$ system[25-27]. They hold the opinion that the dissolution of silicon was a two-step chemical mechanism that can be ascribed to (i) the oxidation of silicon with $HNO_3$ and (ii) the dissolution of $SiO_2$ by HF. According to Turner, $HNO_3$ was the oxidant[28]. Current researches reveal that this model cannot describe the actual reaction process. The product of $HNO_3$ was not only NO but maybe $NO_2$, $N_2O_3$[29-30]. A. Krist[31] et.al concluded that a different oxidation mechanism operated on the SiGe and Si surfaces. J. Kato[32] et.al summarized that the selectivity between SiGe and Si was due to the band offset for the valence-band ($\Delta E_v$ is ~0.12 eV between p+ Si and $Si_{0.7}Ge_{0.3}$). The electrochemical nature of the SiGe/Si etching system could be described as followed: $HNO_3$ got electrons at Cathode and released holes to solution then formed perhaps NO, $NO_2$, $N_2O_3$ according to $HNO_3$ concentration[33]. Conversely, SiGe received these holes to form $Ge^{2+}$ at anode. There exists a chemical equilibrium.

The $Si_{0.7}Ge_{0.3}$ selective etching morphologies with HNA measured with SEM were shown in Figure 1. HNA could get very high selectivity. However, the etching morphology was inhomogeneous, because HNA system was sensitive to dislocation density and delineate defects[18-19]. The bad interface in tunnels was related to epitaxy defects or strain. While at the edge, the interface was relatively smooth because defects moved to the outside surface. Meanwhile, the large relative etching rate between $Si_{0.7}Ge_{0.3}$ and p+ Si was 2.5 ~ 4 nm/s. The large etching rate and time-etching characteristic of HNA made it difficult to precisely control the tunnel depths. Therefore HNA may not be a good method to selectively etch SiGe in small devices.

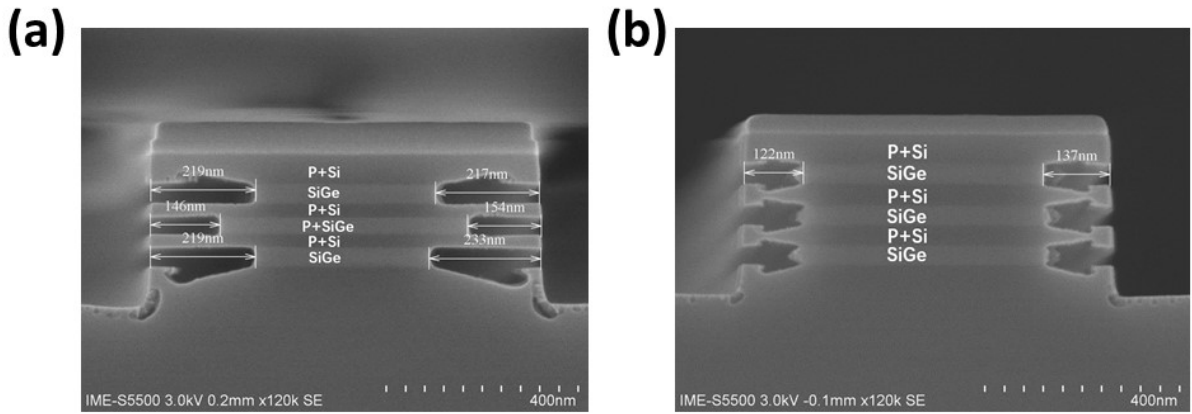

**Figure 1.** SEM images of laminated $Si/Si_{0.7}Ge_{0.3}$ after HNA treatment. (a) Etching time for 60 s. (b) Etching time for 30s. The volume ratio of HF (48%)/ $HNO_3$ (70%)/ $H_2O$ was 10 ml: 784 ml: 716 ml.

The selectively digital etching method is based on the ALE concept[34] and HNA oxidation-etching theory. $HNO_3$ oxidizes Si/SiGe to form oxide and then HF or BOE to remove the oxide. This method was firstly proposed to selectively etch SiGe with precise control and good repeatability. By tuning different oxidation time, we could obtain different selectivity. Moreover, the tunnels with digital etching had no defect pits compared with HNA etching.

3. EXPERIMENTAL SECTION

Our experiments used three types of structures. The starting substrate was 8-inch p-doped (100) bulk Si wafers, then Si and/or SiGe film was epitaxially grown at 650 °C 20 Torr in $H_2$ atmosphere using reduced pressure chemical vapor deposition (RPCVD). $SiH_2Cl_2$ (DCS) and $GeH_4$ were the gaseous precursors of the SiGe layers, while $SiH_4$ was the precursor of the Si layers. Figure 2a was 60 nm intrinsic $Si_{0.7}Ge_{0.3}$, Figure 2b was 110 nm heavily boron doped silicon with concentration of 1E20 $cm^{-3}$, inserting $Si_{0.7}Ge_{0.3}$ for distinguishing with Si substrate, and Figure 2c was 40 nm intrinsic $Si_{0.7}Ge_{0.3}$ and 40 nm p+ Si alternating grown three cycles. Flat $Si_{0.7}Ge_{0.3}$ sample in Figure 2a and flat heavily boron doped silicon (p+ Si) sample in Figure 2b were used to measure the etched amount per cycle (EPC) of flat $Si_{0.7}Ge_{0.3}$ and p+ Si; laminated Si/SiGe in Figure 2c was used to measure the relative etched amount per cycle (REPC) and etching selectivity. EPC was obtained by the total etched amount divided by the total etching cycles, and REPC was obtained by the relative total etched amount, namely tunnel depth, divided by the total etching cycles. The flow diagram was shown in Figure 2d. Before experiment, all samples were immersed in BOE for 3 min to remove the natural oxide. Every experiment contains 60 cycles to reduce the influence of natural oxide thickness and measurement error.

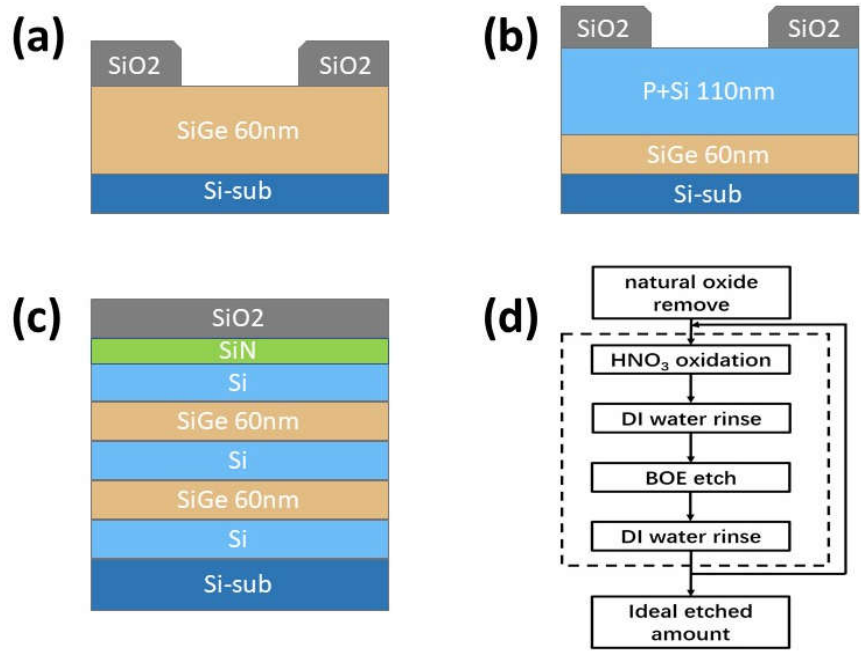

**Figure 2.** Schematics of the sample structures used in our experiments, (a) flat $Si_{0.7}Ge_{0.3}$, (b) flat p+ Si and (c) laminated $Si_{0.7}Ge_{0.3}$/ p+ Si samples. The $SiO_2$ and $Si_3N_4$ stack were the hard mask. (d) The flow chart of $HNO_3$/BOE digital etching.

The $HNO_3$ solutions were diluted with deionized water, and the volumes were the same (2.5 L). By altering the volume ratio of $HNO_3$ (70%) and deionized water to adjust the concentration of $HNO_3$. Because $HNO_3$ solution was easily volatile, monitoring samples were set to ensure experiments stable. Taking sample I 1-30 cycles, sample II 31-60 cycles, and sample III overall 1-60 cycles, the tunnel depths of the 60-cycle sample were found to be twice as much as the 30-cycle samples, and the two 30-cycle samples were the same. This method can also demonstrate that the depth of the tunnels was in proportion to etching cycles.

All the samples were etched by BOE diluted fifty folds with deionized water, and total BOE volume was kept 2 L. When HF concentration was too high, Si or SiGe would be damaged[14, 35]. The etching time was 1 min in our experiments.

The rinsing between the oxidation and etching processes is important. To monitor the effects of the cross-contamination between HNO$_3$ and BOE, SiGe films were immersed in HNO$_3$ and BOE during the experiments. The SiGe in the SiGe films would be etched away if cross-contamination existed. The rinsing time was 1 min.

The shapes of the samples and the etched amount were examined by Scanning Electron Microscope (SEM) from HITACHI and High Resolution Transmission Electron Microscope (HRTEM). Surface roughness and element analysis were measured by Atomic Force Microscope (AFM) from BRUKER and Electron Energy Loss Spectroscopy (EELS).

4. RESULTS AND DISCUSSION

**4.1. Si$_{0.7}$Ge$_{0.3}$ and p+ Si Samples with Flat Surfaces:** We firstly explored the HNO$_3$-BOE selectively digital etching characteristics with flat Si$_{0.7}$Ge$_{0.3}$ and p+ Si films. The etching rate was expressed with EPC and the cycle number was 60. The etched thickness of exposed Si$_{0.7}$Ge$_{0.3}$ was calculated by the thickness under the SiO$_2$ subtracting the remaining exposed Si$_{0.7}$Ge$_{0.3}$ thickness after etch, so as the flat p+ Si. The SEM images of flat Si$_{0.7}$Ge$_{0.3}$ after 80 s oxidation 60 cycles digital etch was shown in Figure 3.

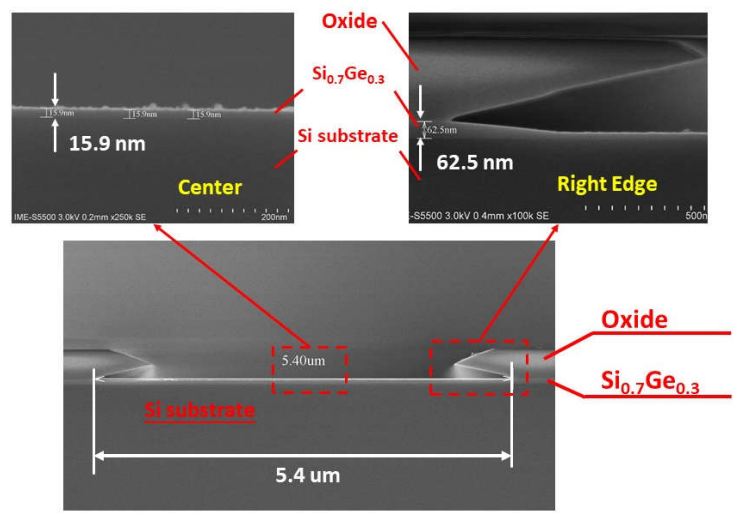

**Figure 3.** The SEM images of flat $Si_{0.7}Ge_{0.3}$ after 80 s oxidation 60 cycles digital etch. The insets are the enlarged figures of the flat surface in the center of the measurement trench and right side of the measurement trench. Etched $Si_{0.7}Ge_{0.3}$ depth was the difference of the thickness under the oxide and the remaining exposed $Si_{0.7}Ge_{0.3}$.

With different oxidation times, both $Si_{0.7}Ge_{0.3}$ and p+ Si were etched. The saturation time and etching rates were 60 s, 0.76 nm/cycle for $Si_{0.7}Ge_{0.3}$ and 120 s, 0.45 nm/cycle for p+ Si with 25% $HNO_3$, as shown in Figure 4.

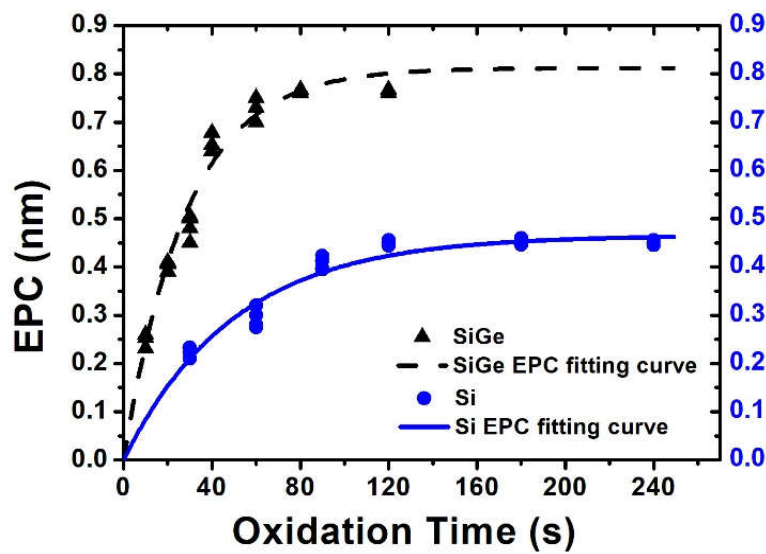

**Figure 4.** The EPCs for $Si_{0.7}Ge_{0.3}$ and p+ Si as a function of oxidation time at $HNO_3$ concentration of 25% and 60 etching cycles of the digital etch. The triangles and the dashed line are from our experiments and our model, shown by eq 3, for $Si_{0.7}Ge_{0.3}$. The circles and the solid line are from experiments and the model, shown by eq 4, for p+ Si.

The reaction in $HNO_3$ was considered as a surface-oxidation reaction[36]. The effective activation energies for the oxidizing flat $Si_{0.7}Ge_{0.3}$ and p+ Si were determined to be 0.45 eV and 0.20 eV from an Arrhenius in Figure 5. Diffusion process dominates in $HNO_3$-BOE digital etch with such low activation energies[21, 31, 37]. The higher effective activation energy for oxidizing $Si_{0.7}Ge_{0.3}$ than p+ Si suggests that high selectivity should be available at high temperature.

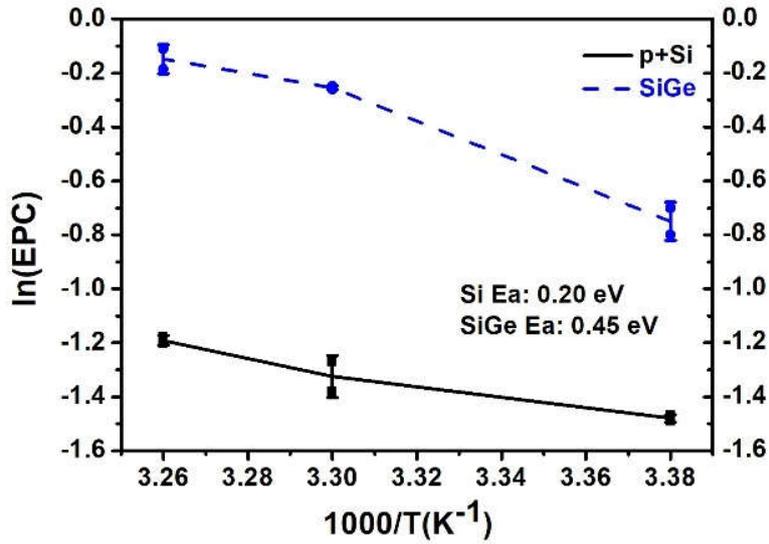

**Figure 5.** Arrhenius plot showing the effective activation energies, 0.45 eV and 0.20 eV, for oxidizing $Si_{0.7}Ge_{0.3}$ and p+ Si respectively.

A physical model was proposed to analyze the oxidation process. As discussed earlier, the oxide thickness growth is a surface reaction and limited by reactant diffusion due to low activation energy at low temperature used in wet oxidation. The oxidation nearly stopped after the oxide thickness reaching a certain value since the oxidant diffusion through the oxide is negligible. Since the surface oxide grew like an island cluster, and eventually converged to a layer. An oxide growth mode was proposed based on this theory, that was

$$\frac{dx(t)}{dt} = \beta \left[1 - \frac{x(t)}{x_0}\right], \quad (1)$$

where $\beta$ and $x_0$ are the oxidation reaction rate of the surface due to SiGe surface exposed to $HNO_3$ or oxidant and the saturation thickness of the oxide, respectively. In eq 1, $x(t)$ is the oxide thickness at time *t*, and the possibility of the sites available for the oxidation is assumed to be proportional to (1 - $x(t)/x_0$). The solution of eq 1 with initial condition $x(0) = 0$ is expressed as,

$$x(t) = x_0 \left(1 - e^{\frac{-t}{\tau}}\right), \text{where } \tau = \frac{x_0}{\beta}. \tag{2}$$

We assume that the thickness of SiGe or Si etched away per cycle is proportional to the thickness of the oxide. Therefore, the EPCs of the $Si_{0.7}Ge_{0.3}$ and p+ Si vs. oxidation times are only different in the values of $x_0$ in eq 2 and then the expressions of EPCs for the $Si_{0.7}Ge_{0.3}$ and p+ Si were obtained by fitting the experimental data in Figure 4,

$$x(t)_{-Si_{0.7}Ge_{0.3}} = 0.81 \left(1 - e^{-\frac{t}{28.3}}\right), \tag{3}$$

$$x(t)_{-p+Si} = 0.47 \left(1 - e^{-\frac{t}{50.0}}\right). \tag{4}$$

We used R-square to measure how successful the fit is in explaining the variation of the data. The R-square is also called the square of the multiple correlation coefficient and the coefficient of multiple determination. R-square can take on any value between 0 and 1, with a value closer to 1 indicating that a greater proportion of variance is accounted for by the model. And the R-squares of $x(t)_{-Si_{0.7}Ge_{0.3}}$ and $x(t)_{-p+Si}$ are 0.95 and 0.93. From eq 3, the EPC of $Si_{0.7}Ge_{0.3}$ is saturated at 0.81 nm (7 ML) and its saturation time is defined as 56.6 s (= $2\tau_{SiGe}$ = 2*28.3). Similarly, the EPC of p+ Si is saturated at 0.47 nm (4 ML) and the corresponding saturation time was 100.0 s from eq 4. The experimental data of EPCs can be described very well, as shown in Figure 4, which indicates the assumption of our model is reasonable for wet oxidation with $HNO_3$. Further experiments are needed to verify if our model can be applied to other wet oxidations.

**4.2. $Si_{0.7}Ge_{0.3}$ and p+ Si Laminated Structures:** By varying oxidation time using 25% $HNO_3$ with 10 s, 30 s, 60 s, 90 s, and 120 s, the samples' sections were shown in Figure 6. The better morphology with $HNO_3$-BOE digital etch than HNA was assumed that oxidation was not as sensitive to defects as etching, oxidation would saturate in a short time and BOE etch time was not long enough to etch Si through defects. So HNO$_3$-BOE digital etch has advantages than HNA etching system for small size devices and heterojunction. From Figure 6, the tunnel depths were almost unchanged when oxidation time was longer than 30 s, and the vertical Si etching depth, namely Si loss, was unchanged when oxidation time was longer than 60 s. It is worth mentioning that when oxidation time was less than 10 s, the tunnel depth evidently reduced, and the etching surface was rough. This was perhaps due to the SiGe/Si oxidation pre-aging time[33] and the faster oxidation rate at surface at the very early stage of oxidation.

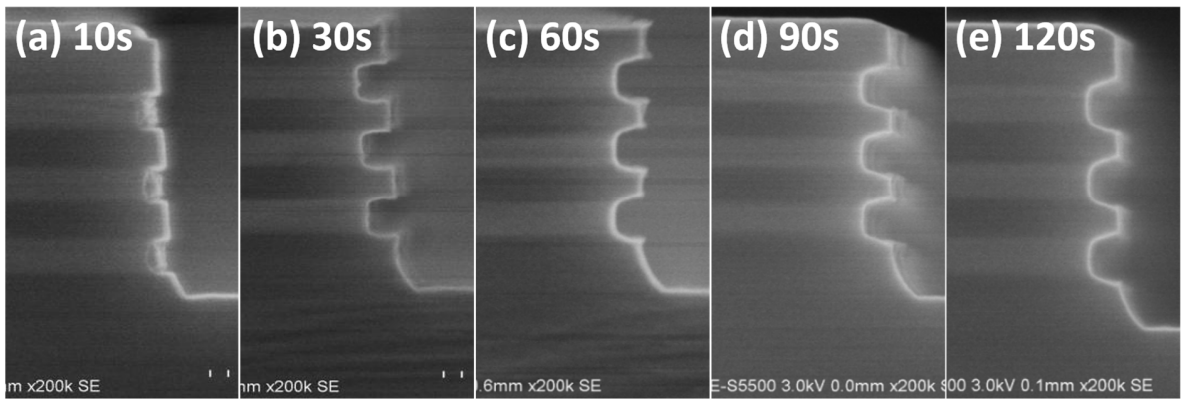

**Figure 6.** The SEM images of laminated p+ Si/Si$_{0.7}$Ge$_{0.3}$ samples after digital etch cycles of 60 at 25% HNO$_3$ with several oxidation times: (a) 10 s, (b) 30 s, (c) 60 s, (d) 90 s and (e) 120 s.

Due to our epitaxy process, Si$_{0.7}$Ge$_{0.3}$ layer upper interface quality was worse than the lower one[11], and the Si loss included the remove of defects. As a result, the upper Si loss was more than the lower one. To verify that digital etch was a surface reaction, elemental analysis with EELS was shown in Figure 7. Sample-1 was the origin sample immersed in 25% HNO$_3$ 24 h and then etched 5 min by BOE, sample-2 was after 25% HNO$_3$ oxidation time 30 s of the digital etch. EELS analysis showed that sample-1 had no obviously relative etching, and elements such as carbon and oxygen exist only at surface, no oxygen atom was found inside SiGe. Above all proved oxidation occurred only at surface and saturated with increasing time. Meanwhile, from HRTEM the sample-2 Si loss was ignorable, which proved the large selectivity.

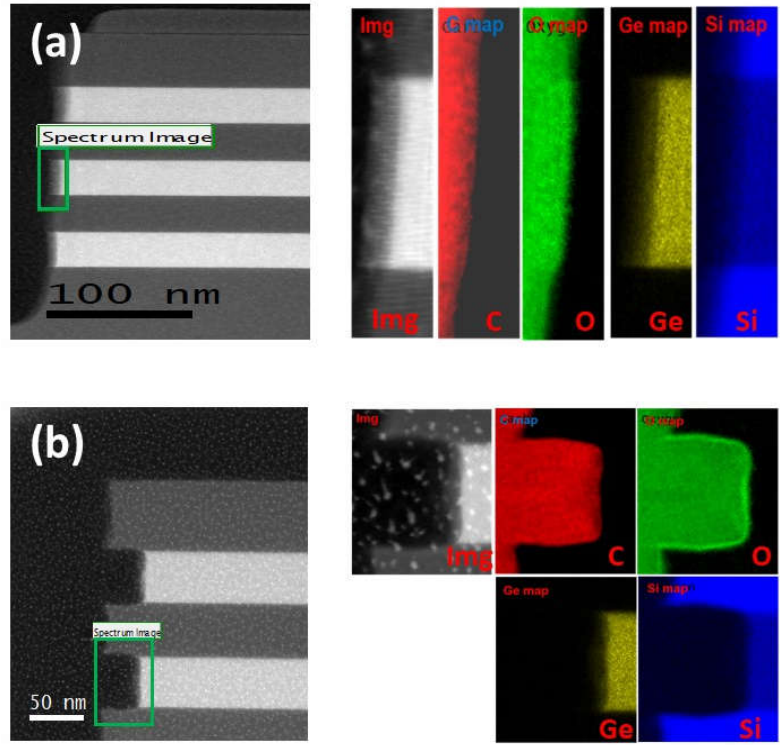

**Figure 7.** HRTEM images and EELS elements analysis of $Si_{0.7}Ge_{0.3}$/p+ Si samples. (a) Sample-1 immersed in 25% $HNO_3$ 24 h, no oxygen atoms were found in the body which proved the oxidation reaction was at surface. (b) Sample-2 experimented by digital etch of 25% $HNO_3$ 30 s 60 cycles.

Selectivity is an important characteristic of selective etching which is the etching rate ratio of desirably etched and unetched materials, and it is preferably high. As shown in Figure 8a, the actual selective etch process was different from ideal process due to not infinite selectivity. According to IBM group[38], selectivity was defined as

$$Selectivity = \frac{Tunnel\ depth}{Si\ loss}. \qquad (5)$$

This formula is accurate when the selectivity is large enough. However, when the selectivity becomes small, the error of the value given by eq 5 is large. For example, for the condition when the angle between the tunnel upper/lower surface and the Si/SiGe interface is 45 degrees, the selectivity is equal to one by eq 5. In this case, however, the two materials, Si and SiGe, should be etched at the same rate, which means the tunnel would not exist or the tunnel depth equals to zero. Therefore, a corrected formula for the selectivity calculation, based on Si and SiGe oxidation-etch path tracing, was proposed and shown in Figure 8a and eq 6. Assumed that the dashed line was the initial envelope lines of the Si and SiGe films of a sample to be etched, and the red circle, Point-1 (P-1), is the initial position of the lower right corner of the top Si film. After a certain etching time *t*, the corner moves from P-1 to Point-2 (P-2), the diamond mark in Figure 8a. The solid lines are the final envelope lines. The displacement vector of the corner could be decomposed into the vertical and horizontal vectors as shown in Figure 8a. The vertical vector or vertical Si loss ($Si_{v\_loss}$) could be measured from SEM or TEM. SiGe was etched horizontally only. The sum of horizontal Si loss ($Si_{h\_loss}$) and tunnel depth was the total etched depth of the SiGe. The selectivity is defined as

$$Selectivity = \frac{Si_{h\_loss} + Tunnel\ depth}{Si_{v\_loss}}. \qquad (6)$$

When vertical Si loss is ignored or tunnel depth is much greater than $Si_{v\_loss}$, the selectivity is large. When both Si and SiGe are etched away but the tunnel depth is almost zero, the selectivity is roughly equal to the ratio between $Si_{h\_loss}$ and $Si_{v\_loss}$, which solve the problem of the selectivity definition given by eq 5. However, we need to find the value of $Si_{h\_loss}$ and $Si_{v\_loss}$. If the influence of crystal planes on the etching rate was ignored, namely $Si_{v\_loss} \approx Si_{h\_loss}$, the selectivity could be defined as

$$Selectivity = \frac{Si_{v\_loss} + Tunnel\ depth}{Si_{v\_loss}}. \qquad (7)$$

Selectivity of the laminated samples with different oxidation time was calculated using eqs 5 and 7 and plotted in Figure 8b. From our experiment, the selectivity was decreasing with time and eventually reached its minimum value of 3.65.

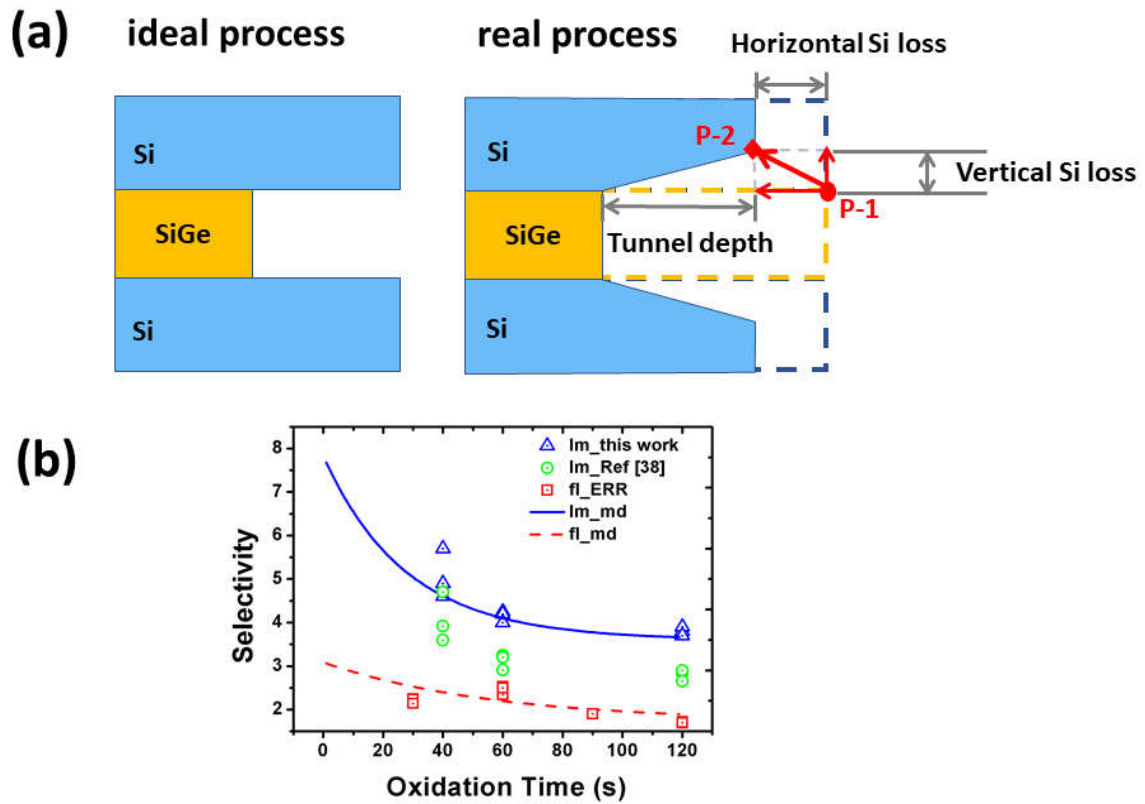

**Figure 8.** (a) 2D schematic diagram of ideal and real process, and selectivity definition method of eq 6. (b) Selectivity as a function of oxidation time. The open triangles and circles are selectivity calculated by our work in eq 7 and Ref 38 in eq 5 with laminated structure (lm) from our experiments, the open squares are etching rate ratios (ERR) of flat $Si_{0.7}Ge_{0.3}$ and p+ Si EPCs (fl) from experiments, the solid line is the selectivity curve calculated from oxidation model (md) of laminated structure by eq 9, and the dashed line is selectivity curve calculated from oxidation model of flat structure by eq 8.

REPC was assumed as the EPC difference of SiGe and Si. From the flat p+ Si and $Si_{0.7}Ge_{0.3}$ oxidation models, REPC of flat structure ($REPC_{fl\_md}$) was described by

$$REPC_{fl\_md} = 0.81\left(1 - e^{-\frac{t}{28.3}}\right) - 0.47\left(1 - e^{-\frac{t}{50.0}}\right). \quad (8)$$

However, the eq 8 cannot fit experiment data well. Consequently, $Si_{0.7}Ge_{0.3}$ and p+ Si parameters should be corrected. From Figure 6, the saturated Si loss was around 12.00 nm, corresponding to saturated oxide thickness of 0.20 nm per cycle, and saturation time was about 60.0 s. Substitute $\tau_{Si}$ = 30.0 s, $x_{0,Si}$ = 0.20 nm and fit experiment data with eq 8, then the corrected $\tau_{SiGe}$ of 13.8 s and $x_{0,SiGe}$ of 0.72 nm were obtained. REPC of laminated structure ($REPC_{lm\_md}$) was shown in eq 9, and the R-square of eq 9 was 0.90. The two fitting curves of eqs 8 and 9 were plotted in Figure 9 together with experiment data.

$$REPC_{lm\_md} = 0.72\left(1 - e^{-\frac{t}{13.8}}\right) - 0.20\left(1 - e^{-\frac{t}{30.0}}\right). \quad (9)$$

From eq 9, $Si_{0.7}Ge_{0.3}$ saturation time was 27.6 s, and saturation oxide thickness was 0.72 nm. $REPC_{lm\_md}$ was initially increasing with time, then slightly decreased, and eventually approached a constant of 0.52 nm (4 ML). Both $Si_{0.7}Ge_{0.3}$ and p+ Si saturation time reduced by half respect to flat condition, however, p+ Si saturation time was nearly twice than $Si_{0.7}Ge_{0.3}$ in both REPC calculation methods. Selectivity was calculated based on our oxidation model, and plotted in Figure 8b. It can be seen that our model fit the experiment well. And the saturated selectivity was 3.6 which was almost the same with the selectivity from the correct selectivity formula (3.65) in eq 7. The highest selectivity of 7.8 can be deduced from Figure 8b. The $REPC_{fl\_md}$ and etching rate ratio (ERR) of $Si_{0.7}Ge_{0.3}$ and p+ Si were also plotted in Figure 9 and Figure 8b. The calculated results from two flat surface samples were lower than experiment results. The two different consequences were found which was maybe due to heterojunction strain, defects and/or changed electrochemical reaction.

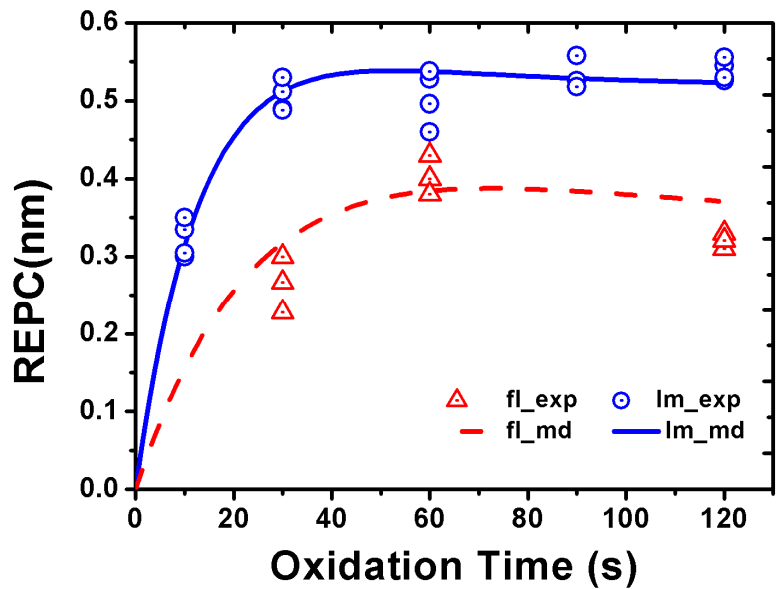

**Figure 9.** The REPCs as a function of oxidation time at $HNO_3$ concentration of 25% and 60 cycles of the digital etch. The open circles and the solid line were the REPCs from our experiments (exp) and model ($REPC_{lm\_md}$ in eq 9), for laminated structures. The open triangles and the dashed line were the REPCs from experiments and model ($REPC_{fl\_md}$ in eq 8), for flat surface samples.

CONCLUSIONS

In this work, $HNO_3/HF/H_2O$ etching silicon theory has been verified by separated oxidation and etching. And $HNO_3$-BOE digital etch was proposed based on this theory. It was found that oxidation would saturate with time. An oxidation model was proposed to describe this oxidation process and selective etching. From the model, the oxidation saturated time and saturated thickness were 100.0 s and 0.47 nm for flat p+ Si, 56.6 s and 0.81 nm for flat $Si_{0.7}Ge_{0.3}$ of 25% $HNO_3$ 60 cycles of digital etch. While in laminated structure, the oxidation saturated time and saturated thickness were 27.6 s and 0.72 nm for $Si_{0.7}Ge_{0.3}$, 60.0 s and 0.20 nm for p+ Si. The saturated selectivity of $Si_{0.7}Ge_{0.3}$/p+ Si stacks was 3.6 and variation was 4% with 30.8% $HNO_3$. The model fit well with our experiments. A corrected selectivity calculation formula was proposed, and the selectivity from the formula was the same with selectivity from the model. The digital etch method could be used to obtain controllable and accurate Si/SiGe tunnel depth and the saturated REPC was about 4 ML.

ACKNOWLEDGMENT

The authors would like to thank team members for the accomplishment of dozens of times experiments. This research was funded by the Academy of Integrated Circuit of the Chinese Academy of Sciences under Grant Y7YC01X001.